\documentstyle[preprint,revtex]{aps}
\begin{document}
\preprint{Submitted to Europhys. Lett.}
\draft
\begin{title}
Universal Conductance Distribution in the Quantum Size Regime.
\end{title}
\author{Alex Kamenev and Yuval Gefen}
\begin{instit}
Department of Condensed Matter Physics, The Weizmann Institute of Science,
Rehovot 76100, Israel
\end{instit}
\begin{abstract}
We study the  conductance ($g$) distribution function of an ensemble of
isolated conducting rings, with an Aharonov--Bohm flux. This is done in
the discrete spectrum limit, i.e., when the inelastic rate, frequency and
temperature are  all smaller than the mean level spacing. Over a wide range
of $g$  the distribution function exhibits universal behavior
$P(g)\sim g^{-(4+\beta)/3}$, where $\beta=1 (2)$ for  systems with (without)
a time reversal symmetry. The nonuniversal large $g$ tail of this
distribution determines the values of high moments.
\end{abstract}


The conductance of weakly disordered (diffusive) mesoscopic systems is not a
self--averaging quantity and exhibits enormously large
sample-to-sample fluctuations \cite{Stone85}. It has been shown
\cite{Altshuler88} that for a good metal the conductance
distribution function (CDF) is nearly a Gaussian  centered around the
average (Drude) conductance \mbox{$\langle g\rangle \approx g_0=E_c$},
where $E_c$ is a Thouless
correlation energy (hereafter we take  $e^2/\hbar =1$ and
the mean level spacing  $\Delta= 1$).
The width of the distribution is disorder independent (but weakly geometry
dependent)
$\langle \delta g^2\rangle\equiv
\langle g^2\rangle-\langle g\rangle^2\approx b\beta^{-1}$, where $\beta=1,2$
in the orthogonal ensemble (GOE) and in the unitary ensemble (GUE)
respectively, $b$ being of order $1$.
These results  were obtained for mesoscopic granules attached
to conductive leads. They rely on the picture that after  diffusing
throughout the system during time $\sim\hbar/E_c$, an electron is absorbed
(inelastically scattered) by the leads.
The associated uncertainty in the position of a level (level broadening) is
$\gamma\approx E_c \gg 1$.
{}From the technical point of view the presence of leads enables
one to employ either  diagrammatic perturbation theory (without the zero
mode)
\cite{Stone85}  or a Landauer type approach \cite{Stone89}.

The question at hand is how the CDF is affected if $\gamma$ is made smaller
than $E_c$, and eventually smaller than $1$ (i.e., the discrete spectrum
limit, where   the interlevel spacing exceeds the level broadening).
This condition is achieved when the coupling of
the system to the external  reservoir (leads)
(e.g. through the tunneling barriers),  is made weak or is absent in the
first place.  We note that
the origin of level broadening may be due to  coupling with an energy (rather
than a particle)
reservoir (e.g. phonons). Perturbative calculations,
employed in the range $1\ll\gamma$, yield (for the zero dimensional
case) \cite{Stone85}
\begin{equation}
\langle \delta g^2\rangle\approx
\frac{\tilde b}{\beta}\left(\frac{E_c}{\gamma}\right)^2=
\frac{\tilde b}{\beta}\frac{g_0^2}{\gamma^2}.
                                                           \label{pert}
\end{equation}
This implies that the width of the Gaussian CDF increases as $\gamma$
decreases. Equation (\ref{pert}) is not valid for
$\gamma\leq 1$ as perturbation theory breaks down at $\gamma\approx 1$.

In this paper we focus on the CDF in the  quantum size regime.
By this we
imply that the inelastic rate, the external frequency and the temperature are
all smaller  than
the mean level spacing: $\delta\equiv \max\{\gamma,\omega,T\}\ll 1$.
A similar problem has been recently considered by
Prigodin, Efetov and Iida  \cite{Efetov93}, who used the Landauer formula
to calculate the
conductance of a disordered metallic grain placed between two point-like
contacts.  For $\gamma \ll 1$ and the GUE they
obtained the following behavior of the CDF  \cite{Efetov93}
$$P(g)\propto \left(\frac{g}{\langle g\rangle}\right)^{-3/2}; \,\,\,
\gamma\langle g\rangle\ll g \ll \langle g\rangle/\gamma. $$
For smaller values of $g$ the CDF, $P(g)$, saturates to a constant, while
for larger $g$ it decays exponentially.
The result of Ref.\ \cite{Efetov93} is  valid for  noninteracting electrons,
studied within the grand canonical ensemble (GCE).
However, these authors stress \cite{Efetov93} that for a sample with two
point-like contacts  interaction effects (i.e. Coulomb blockade) play a very
important role. This is why their CDF is not amenable to experimental
testing.

Here we focus on a different  setup, which is clearly advantageous from the
theoretical (and possibly from the experimental) point of view.
Consider a large ensemble of metallic
rings each threaded by a  time dependent Aharonov--Bohm flux,
$\Phi(t)=\overline\Phi+\delta\Phi(t)$. The linear dissipative conductance
of a ring is defined as the ratio between the current flowing around the ring
and the electromotive force:
$$g(\overline\Phi,\omega)=\Re\frac{I(\overline\Phi,\omega)}
{i\omega\delta\Phi(\omega)}.$$
We shall be interested in the distribution function of $g$.
The obvious merits of the Aharonov--Bohm geometry are
\begin{itemize}
\item As there are no contacts attached,  the level broadening
$\gamma$ may be made arbitrarily small for sufficiently small temperature.
\item There is practically no  screening of the external electric field
within the volume of the
conductor (the skin radius is much larger than the
crossection of the ring). This means that the large wavelength internal field
is equal to the  external field \cite{Landauer92}.
\item The number of particles in each ring is fixed (canonical statistical
ensemble). For this reason the
issue of Coulomb blockade does not arise.
\end{itemize}
Under canonical conditions and for $T\ll 1$ (necessary to suppress
intralevel transitions which, otherwise, present at
$\overline\Phi\neq 0$) the sample specific conductance of the ring is given by
\cite{Trivedi88,Kamenev94}.
\begin{equation}
g(\overline\Phi,\omega)=\sum_{n\neq m}
\frac{f_m-f_n}{\epsilon_n-\epsilon_m}
\frac{\gamma }{\gamma^2+(\epsilon_n-\epsilon_m-\omega)^2}|I_{n,m}|^2
                                                            \label{cond}
\end{equation}
where $\{\epsilon_n(\overline\Phi)\}$ and  $\{|n(\overline\Phi)>\}$ are the
exact eigenvalues and eigenfunctions of the single electron Hamiltonian,
$f_n\equiv f(\epsilon_n(\overline\Phi)-\mu(\overline\Phi))$ is the  Fermi
function and \mbox{$I_{n,m}\equiv<n|\hat I|m>$} is the matrix element of the
current operator. In the quantum size regime the
largest contribution to the double sum in Eq. (\ref{cond})
comes from the  pair of consecutive levels situated on opposite sides of the
Fermi energy, $\epsilon_m<\epsilon_F<\epsilon_{m+1}$.
Restricting ourselves to this particular pair  (below we shall discuss
the range of validity for such an approximation) we obtain
\begin{equation}
g\approx
\frac{f_m-f_{m+1}}{\epsilon}\left[
\frac{\gamma}{\gamma^2+(\epsilon-\omega)^2}+
\frac{\gamma}{\gamma^2+(\epsilon+\omega)^2} \right] |I|^2,
                                                            \label{cond1}
\end{equation}
where $\epsilon=\epsilon_{m+1}-\epsilon_m$ and $I=I_{m+1,m}$ are random
quantities depending on the realization of the impurity potential.
Employing the result of random matrix theory that the statistics of the
eigenenergies and the eigenvectors of a random Hamiltonian are independent
(within the GOE and GUE limits, but not necessarily within intermediate
ensemble) \cite{Mehta91},
it follows that $\epsilon$ and $|I|$ are
statistically independent as well. The distribution function of the
nearest level spacing is well approximated by the Wigner surmise
\cite{Mehta91}
\begin{equation}
P(\epsilon)=a_{\beta}\epsilon^{\beta}
\exp\left[-b_{\beta}\epsilon^2\right],
                                                            \label{wigner}
\end{equation}
where $a_1=\pi/2,\, a_2=32/\pi^2;\, \, \, \, b_1=\pi/4,\, b_2=4/\pi$.
We stress that we employ the canonical statistical ensemble,  meaning
that the number of particles in each system is a constant assigned to each
ring  independently of the
its specific impurity configuration. In other words, for each ring the last
occupied level (at $T=0$) is chosen at random.
Thus the energy gap, $\epsilon$, situated
near the Fermi energy, is selected at random (from all possible values of the
gap) and it's distribution is given by Eq.\ (\ref{wigner}), \cite{foot2}.

The matrix element $|I|$ may be shown \cite{French88} to have a Gaussian
distribution function, which may be written as
\begin{equation}
P(|I|)=\frac{2}{\sqrt{2\pi\langle |I|^2 \rangle }}
\exp \left[  -\frac{|I|^2}{2\langle |I|^2 \rangle} \right].
                                                            \label{gauss}
\end{equation}
The  simplest way to establish the value of the single parameter of this
distribution, $\langle |I|^2 \rangle$, is to evaluate  the average
conductance from Eq.\ (\ref{cond}) for  $\gamma\gg 1$, \cite{Kamenev94}.
The result
should be equal (up to small flux dependent corrections) to the Drude
conductance, $g_0=E_c$. We thus  obtain
$$\langle |I|^2 \rangle=g_0/\pi.$$

The problem now is to calculate the distribution function of $g$ expressed by
Eq.\ (\ref{cond1}), given the  distributions for $\epsilon$ and $|I|$,
Eqs.\ (\ref{wigner}) and (\ref{gauss}). We first comment on the range of
validity
of the two level approximation, Eq.\ (\ref{cond1}). Suppose that the energy
gap near the Fermi energy is of the order of the mean level spacing,
$\epsilon\approx 1$. In this
case the contributions to $g$ of the levels next to the pair at hand is of the
same order of magnitude ($\approx\gamma |I|^2\approx \gamma g_0$) as the
original
pair. (Note that after summation over all pairs of levels the order of
magnitude of the
conductance remains the same due to the fast convergence of the sum ---
the respective contribution of the $(m,n)$ pair falls of as
$(\epsilon_n-\epsilon_m)^{-3}$).
For such a relatively large gap the probability to find
values of the conductance
much larger than $\gamma g_0$ is associated with large
fluctuations  in the  matrix elements, $|I|$, and is therefore  exponentially
small (cf. Eq.\ (\ref{gauss})). At the same time
the probability to find
the conductance much smaller than $\gamma g_0$ is exponentially small as
well.
Indeed, to obtain a value of the conductance which is nearly zero we require
{\em all} $|I_{m,n}|\approx 0$. The probability for the latter vanishes
faster
than a power law. We thus conclude that all  systems with the
energy gap, $\epsilon$, being of the order or larger than the mean level
spacing yield  conductance of the
order of $g\approx \gamma g_0\ll g_0$. The probability to find the conductance
of such a system $g\gg \gamma g_0$ or $g\approx 0$ is exponentially small.
By contrast,  systems with $\epsilon\ll 1$ contribute to
the large $g$ ($g\gg\gamma g_0$) part of the CDF; in this case the two level
approximation, Eq.\ (\ref{cond1}), is applicable.

For  systems with $\delta\ll\epsilon\ll 1$, employing the two level
approximation, the conductance may be written
as (cf. Eq.\ (\ref{cond1})) $g=2\gamma|I|^2/\epsilon^3$.
The CDF is given by
\begin{equation}
P(g)\approx\int_0^{\infty} d|I| P(|I|)\int_{\delta}^{1}d\epsilon P(\epsilon)
\delta(g-2\gamma|I|^2/\epsilon^3) .
                                                            \label{formal}
\end{equation}
After  integration one obtains
\begin{equation}
P(g)\approx c_{\beta} (\gamma g_0)^{\frac{1+\beta}{3}}
g^{-\frac{4+\beta}{3}};\, \, \, \, \,
\gamma g_0\ll g \ll \gamma g_0/\delta^3,
                                                            \label{main}
\end{equation}
where
$$c_1=\frac{1}{18}\left( \frac{2}{\sqrt{\pi}} \right)^{1/3}
\Gamma\left(\frac{1}{6}\right), \, \, \,
c_2=\frac{64}{3\pi^3}.$$
The second inequality in Eq.\ (\ref{main}) is discussed bellow.
Eq.\ (\ref{main}) is our main result. It
shows that the bulk of the CDF exhibits a power law
behavior: the distribution function decreases as $g^{-5/3}$ in the GOE and as
$g^{-2}$ in the GUE.
This should be compared with $g^{-3/2}$ in the GUE for the systems with
point--like contacts \cite{Efetov93,foot3}.
The bulk of the CDF,
$\gamma g_0\ll g \ll \gamma g_0/\delta^3$, is associated with  realizations
satisfying
$\delta\ll\epsilon\ll 1$. As we argued above, systems with
$\epsilon\geq 1$ contribute only to the $g\leq \gamma g_0$ part of the CDF and
introduce only exponentially small corrections to the $g>\gamma g_0$ region.
Systems with $\epsilon\leq\delta\ll 1$ may contribute, in principle,
to the bulk of the CDF, provided $|I|^2\ll\langle |I|^2\rangle$.
However, the relative fraction of such
realizations is $\delta^{1+\beta}\ll 1$, thus they do not change the
character of the body of the CDF. Otherwise, these are precisely these
rare realizations
($\epsilon\leq\delta\ll 1$) which determine the large $g$
($g\geq\gamma g_0/\delta^3$)
tail of the CDF. Table. 1 depicts schematically the various regimes of
$\epsilon$, which determine the different ranges of $P(g)$.
All  moments of the CDF, starting from the
second one, are determined by the large $g$ tail. Thus high moments of the
distribution function carry information about a small number of atypical
realizations with very small gaps near the Fermi energy.
The form of the tail is nonuniversal and depends on the
relations among the inelastic rate, the frequency and the temperature.
We demonstrate the above by studying one example.

Let us consider the case  $\omega=0;\, T\ll \gamma\equiv \delta\ll 1$.
Eq.\ (\ref{cond1}) may be then rewritten as
\mbox{$g\approx 2|I|^2/(\gamma\epsilon)$} for $T\ll\epsilon\ll\gamma $ and
\mbox{$g\approx |I|^2/(2\gamma T)$} for $\epsilon\ll T $. Computing
integrals similar to Eq.\ (\ref{formal}) one obtains
\begin{equation}
P(g)\approx\left\{ \begin{array}{ll}
\displaystyle{
d_{\beta}\left( \frac{g_0}{\gamma} \right)^{1+\beta} g^{-(2+\beta)}    };
\hskip 1cm     &
\displaystyle{ \frac{g_0}{\gamma^2}\ll g\ll\frac{g_0}{\gamma T}   }, \\
\displaystyle{
\frac{a_{\beta}}{1+\beta}T^{1+\beta}\sqrt{\frac{\gamma T}{g_0 g}}\,
\exp\left[-\pi\frac{\gamma T g}{g_0}\right]   };
\hskip 1cm     &
\displaystyle{ \frac{g_0}{\gamma T }\ll g   },
\end{array}
\right.
                                                          \label{tail}
\end{equation}
where $d_1=6/\pi,\, d_2=3840/\pi^5$.
The full behavior of the CDF, in this case is depicted schematically in
Fig. 1. Other scenarios, corresponding to different relations among the
various energy scales, may be worked out in a similar fashion.
Summarizing the main features of the CDF
\begin{itemize}
\item the CDF goes to zero faster than any power for $g\ll g_0\gamma$;
\item the main body of the CDF has a universal power--law behavior
$\sim g^{-(4+\beta)/3}$ for \mbox{$g_0\gamma\ll g\ll g_0\gamma\delta^{-3}$};
\item for $g\gg g_0\gamma\delta^{-3}$ there is a range of nonuniversal
power law decay;
\item for large values of $g$ the CDF has an exponential tail.
\end{itemize}
Eqs.\ (\ref{main}),(\ref{tail}) allow  one to estimate the moments of the
conductance.
The average conductance is determined by the universal part of the
CDF, Eq.\ (\ref{main}), and it is given by
\begin{equation}
\langle g\rangle\sim\left\{ \begin{array}{ll}
g_0; \hskip 1.5cm    & \beta=1, \\
g_0\gamma\ln\gamma;   & \beta=2.
\end{array}
\right.
                                                          \label{mean}
\end{equation}
The average conductance in the quantum size regime in the canonical
statistical ensemble
exhibits {\em negative} magneto conductance. This fact has already been
pointed out
in Ref. \cite{Kamenev94}. For the variance of the conductance one obtains
\begin{equation}
\langle \delta g^2 \rangle \sim
\frac{g_0^2}{\gamma^2}\left\{ \begin{array}{ll}
\displaystyle{\ln\frac{\gamma}{T}  }; \hskip 1.5cm    & \beta=1, \\
\gamma;   & \beta=2,
\end{array}
\right.
                                                          \label{var}
\end{equation}
which by and large agrees with the result of perturbation theory,
Eq.\ (\ref{pert}), for
$\gamma\approx 1$. In agreement with the
perturbative regime the CDF is narrower in the unitary ensemble.

It is of interest to consider the crossover from the GOE to the GUE.
This crossover takes place for
$0\leq\overline\Phi\leq\Phi_c=\Phi_0/\sqrt{g_0}$ \cite{Montambaux92}.
In the crossover regime  level statistics correspond to the GUE for small
energy intervals and to the GOE for larger energy separations \cite{French88}.
For  $\overline\Phi/\Phi_c\leq\delta\ll 1$  only the  non--universal large $g$
tail of
the CDF is affected by the magnetic flux, while the universal part of the CDF
remains as in the GOE. For larger values of flux,
$\delta\ll\overline\Phi/\Phi_c\ll 1$,
the universal part of the CDF may be divided into two regimes \cite{foot4}
\begin{equation}
P(g)\sim\left\{ \begin{array}{ll}
\displaystyle{
(\gamma g_0)^{2/3} g^{-5/3} ; }
\hskip 1cm     &
\displaystyle{ \gamma g_0\ll g\ll\gamma g_0
\left( \frac{\Phi_c}{\overline\Phi} \right)^3   }, \\
\displaystyle{
\frac{\Phi_c}{\overline\Phi}(\gamma g_0) g^{-2}; }
\hskip 1cm     &
\displaystyle{\gamma g_0 \left( \frac{\Phi_c}{\overline\Phi} \right)^3 \ll
g \ll \gamma g_0/ \delta^{3}  },
\end{array}
\right.
                                                          \label{cross}
\end{equation}
Finally for $\overline\Phi/\Phi_c\geq 1$ the CDF has practically assumed the
GUE form.

To summarize, we have derived the distribution function of the conductance
for an ensemble of mesoscopic rings threaded
by a Aharonov--Bohm flux, with a fixed number of particles .
When level broadening, temperature and frequency are all smaller than the
mean level spacing, the distribution function is non--Gaussian. The body of
the distribution function, however, is described by a universal power law
behavior, Eq.\ (\ref{main}). At the same time, higher moments of the
conductance are determined by the nonuniversal tail of the distribution,
which depends on the relations among level broadening,
temperature and frequency.

We acknowledge discussions with I. Lerner and V. Prigodin.
This research was supported by the German--Israel
Foundation (GIF) and the U.S.--Israel Binational Science Foundation (BSF).

\begin{table}
\caption
{Various contributions to $P(g)$ comming from different ranges of the energy
gap, $\epsilon$,  near the Fermi energy. }
\end{table}
\begin{tabular}{|c||c|c|c|}
\hline
 & $\epsilon<\delta$   &  $\delta<\epsilon<1$&$ 1<\epsilon $  \\ \hline \hline
\multicolumn{1}{|c||}{$g<\gamma g_0$} &
\multicolumn{3}{p{1.75in}|}{ two levels approximation is
                     not applicable } \\ \hline
$\gamma g_0<g<\gamma g_0 \delta^{-3}$ & $\sim\delta^{1+\beta}$ &
                        Eq.\ (\ref{main}) &          exp small \\ \hline
$\gamma g_0 \delta^{-3} <g          $ & Eq.\ (\ref{tail})  & exp small
&  exp small \\ \hline
\end{tabular}

\figure{The conductance distribution function (schematic);
GOE-- solid line, GUE--dashed line . Here  $\gamma=0.45.$ The
almost Gaussian distribution for systems with leads is depicted by the dotted
line.
}


\begin{references}
\bibitem{foot3}Another important difference
is the behavior of the CDF at small $g$: according to Ref.\ \cite{Efetov93}
the CDF saturates to a constant for $g\ll \gamma \langle g \rangle$,
while in our case it goes to
zero  faster than any power of $g$ as $g\rightarrow 0$.
\bibitem{foot4}To obtain Eq.\ (\ref{cross}) we employed the Wigner surmise in
the crossover regime \cite{French88} and used again the statistical
independence of matrix the matrix elements and eigenenergies (see discussion
after Eq.\ (\ref{cond1})). The last statement can not be proven for the
crossover regime. We, nevertheless, believe that the main features of the CDF
come from the energy gaps statistics and thus Eq.\ (\ref{cross}) is
qualitatively true.
\end{references}
\end{document}